\pdfoutput=1
\documentclass[fleqn,usenatbib]{mnras}
\usepackage{newtxtext,newtxmath}
\usepackage{xcolor}
\usepackage[T1]{fontenc}

\DeclareRobustCommand{\VAN}[3]{#2}
\let\VANthebibliography\thebibliography
\def\thebibliography{\DeclareRobustCommand{\VAN}[3]{##3}\VANthebibliography}

\usepackage{soul}
\usepackage{longtable}
\usepackage{graphicx}	% Including figure files
\usepackage{amsmath}	% Advanced maths commands
\title[]{Can massive stars form in low mass clouds?}
\author[J. D. Smith et al.]{
Jamie D. Smith,$^{1}$\thanks{E-mail: j.smith49@herts.ac.uk (JDS)}
Sarah E. Jaffa$^{1,2}$
and Martin G. H. Krause$^{1}$
\\
$^{1}$Centre for Astrophysics Research, Department of Physics, Astronomy and Mathematics, University of Hertfordshire, College Lane, Hatfield, Hertfordshire AL109AB, UK\\
$^{2}$Advanced Research Computing, University College London, London, WC1H 9NE, UK
}

\date{Accepted XXX. Received YYY; in original form ZZZ}

\pubyear{2023}

\begin{document}
\label{firstpage}
\pagerange{\pageref{firstpage}--\pageref{lastpage}}
\maketitle
%\listoftodos
\begin{abstract}
  The conditions required for massive star formation are debated, particularly whether massive stars must form in conjunction with massive clusters. Some authors have advanced the view that stars of any mass (below the total cluster mass) can form in clusters of any mass with some probability (random sampling). Others pointed out that the scatter in the determinations of the most massive star mass for a given cluster mass was consistent with the measurement error, such that the mass of the most massive star was determined by the total cluster mass (optimal sampling). Here we investigate the relation between cluster mass (M\textsubscript{ecl}) and the maximum stellar mass (M\textsubscript{max}) using a suite of SPH simulations. Varying cloud mass and turbulence random seed results in a range of cluster masses which we compare with their respective maximum star masses. We find that more massive clusters will have, on average, higher mass stars with this trend being steeper at lower cluster masses ($M\textsubscript{max} \propto M\textsubscript{ecl}^{0.31}$ for $M\textsubscript{ecl}<500M\,_{\odot}$) and flattening at higher cluster masses ($M\textsubscript{max} \propto M\textsubscript{ecl}^{0.11}$ for $M\textsubscript{ecl}>500M\,_{\odot}$). This rules out purely stochastic star formation in our simulations. Significant scatter in the maximum masses with identical initial conditions also rules out the possibility that the relation is purely deterministic (that is that a given cluster mass will result in a specific maximum stellar mass). In conclusion our simulations disagree with both random and optimal sampling of the initial mass function.

\end{abstract}


%%%%%%%%%%%%%%%%%%%%%%%%%%%%%%%%%%%%%%%%%%%%%%%%%%

%%%%%%%%%%%%%%%%% BODY OF PAPER %%%%%%%%%%%%%%%%%%
\section{Introduction}
The stellar Initial Mass Function (IMF) is a crucial tool when studying star formation, stellar evolution, and galaxy evolution \citep[e.g.][]{Bastian2010uniIMF,Guszejnov2022,sharda2022bottonheavy,tabassum2022}. Key features of the IMF (e.g. location of the peak, upper mass limit, and slope of the high mass end) are all important indicators when studying the formation of stars and star clusters. There remains ongoing debate as to whether the IMF is universal - that is to say a random sample of stars taken from the IMF would be a legitimate stellar population independently of the various initial conditions (e.g. mass of parent cloud, turbulence, local environment) that affect star formation.
For example, \cite{Weidner2010Mmass_Mecl} studied a set of star clusters and their mass functions and concluded that it is unlikely that random sampling is correct for their data sample.

Studies, both observational \citep[e.g.][]{Andrews2014,Weidner2010Mmass_Mecl} and using simulations \citep[e.g.][]{bonnell2004,Popescu2014montecarlo}, have been performed to ascertain whether there is necessarily a direct link between the mass of a cluster and the mass of its most massive star. One issue in this context is whether there is a fundamental upper limit to stellar masses \citep[e.g.][]{Weidnerkroupa04}. The most massive stars known to date are located in R136 in the Large Magellanic Cloud and are inferred to have had initial masses > $250\,M_{\odot}$ \citep{brands2022}. It is possible that runaway collisions further increase the star masses in massive clusters, possibly even above $1,000\,M_{\odot}$ \citep[e.g.][]{gieles2018}, but it is difficult to obtain observational evidence for such stars \citep[][]{Nowak2022}.

\citet{Weidner2010Mmass_Mecl} conduct an observational study of Milky Way clusters and find that the mass of the most massive star (M\textsubscript{max}) increases with cluster mass (M\textsubscript{ecl}) up to $\sim 120\,M_{\odot}$, with the data suggesting a power-law relation between M\textsubscript{max} and M\textsubscript{ecl}. \citet{Weidner2013MmaxMecl2} then suggested that the scatter in the relation was purely observational uncertainties and that M\textsubscript{max} was fully determined by the cluster mass. They called this an optimally sampled mass function.

\cite{Andrews2014} used the ionising photon flux of a cluster to infer the presence of massive stars. They found clusters with fluxes that would be inconsistent with the predictions of the $M_\mathrm{max}-M_\mathrm{ecl}$ relation, contrary to the predictions of \citet[][]{Weidner2013MmaxMecl2}. While the findings of \cite{Andrews2014} seem conclusive, we note that their findings are based on the inferred presence of massive stars from unresolved clusters, and they note many sources of uncertainty with their data \citep[compare also][]{Weidner2014NGC4214}.

Attempts have also been made to locate massive stars that don't have a nearby cluster that they could have formed in \citep[e.g.][]{bestenlehner2011VLTFLAMES,Bressert2012VLT-FLAMES,chu2008,deWit2004}, though the presence of bow shocks supports the theory that many of these massive field stars are runaways \citep[e.g.][]{deWit2005,gvaramadze2008OBrunaway}. \cite{Oskinova2013Isolation} claim to have found an even more massive star with no bow shock and no obvious parent cluster candidate.

The problem was addressed theoretically by \cite{bonnell2004}. These authors simulated turbulent molecular clouds to evaluate the effect that fragmentation and competitive accretion have on the massive star formation. They find that the final mass of the most massive star is not correlated with the mass of the clump it formed from but instead is dependent on the competitive accretion that results from the continuing cluster formation. 
\cite{bonnell2004} also found a correlation of the most massive star with the mass of the host cluster, measured by taking a subsample of stars around the chosen massive star in a simulation of one turbulent cloud with an initial mass of $1,000\,M_\odot$. 
In this work we are expanding this study in two important ways. First, we present a suite of simulations now spanning a large range of parent cloud masses. Second, we investigate the statistical variations of the results by repeating each simulation multiple times with different random seeds for the generation of the initial turbulent state. We find true scatter in the M\textsubscript{max}-M\textsubscript{ecl} relation in clear contradiction to the deterministic expectations of optimal sampling. We also show that the dependence of our mean most-massive-star mass on cluster mass disagrees with the expectations from random sampling. In a resolution study we demonstrate that some dependence of the most massive star mass on the mass of the host cluster remains even at our highest resolution.

\section{Method}
\label{method}
We performed a series of Smooth Particle Hydrodynamics (SPH) simulations of turbulent, isolated, gravitationally collapsing clouds, sweeping two different parameters. First we varied the initial mass and radius of the cloud to maintain the initial density which is the same for all simulations in the present paper. This initial density is $1.62\times 10^{-22}\mathrm{g/cm}^{3}$. This ensures that the initial free-fall time, $\propto 1/\sqrt{G\rho}$, is the same for all simulations. This justifies a common simulation time and comparison of the states of all simulations at the same time. Second we increased the number of SPH particles by factors of 2 for each of the masses to see how the improvement to mass resolution affected the properties of the sink particle population.

All the simulations were performed using the SPH code GANDALF \citep{Hubber_2017}. We initiated isolated spherical turbulent clouds following \cite{Jaffa2022}. Our simulations are isothermal at $10\,\mathrm{K}$ as it is a good approximation for molecular clouds at the densities that we resolve \citep{Krumholz2015}. The virial parameter is set to 1. We simulated the system's evolution without feedback until $5\,\mathrm{Myr}$ has passed. We choose $5\, \mathrm{Myr}$ for our end time as it is long enough for a decent period of star formation but not too long that the effects of feedback would be too significant. Some simulations don't reach $5\, \mathrm{Myr}$ due to numerical issues such as the timestep becoming too small. Gravitational forces are computed using a KD tree \citep[full description in][]{Hubber_2017}. Sink particles are used to replace gas that surpasses a certain critical density and is not too close to an existing sink. This density criterion is resolution dependent and corresponds to resolving the Jeans mass with at least 100 SPH particles \citep{Bate1997sphresolution}. These sink particles can accrete after they are formed but are not allowed to merge. Sink particles will accrete SPH particles that are within their accretion radius and are gravitationally bound to the sink. All this follows the original setup of \cite{bonnell2004} closely.

Various forms of feedback have been implemented in recent simulations of star formation. They generally damp the growth of stars, with some forms being particularly effective at inhibiting growth at higher star masses \citep{Bastian2010uniIMF,Guszejnov2022,sharda2022bottonheavy,tabassum2022}. In the present paper we neglect any form of feedback as it would only inhibit the formation of massive stars. Our most massive stars already fall short of observed star masses at the high mass end at the chosen end time of our simulations. Additionally, any effect of feedback would have to affect high or low mass clusters differently to significantly change our results.

For the mass variation, we simulate clouds spanning the mass range $1000\,M_{\odot}$ to $40,000\,M_{\odot}$. The radii of the clouds are varied to maintain constant initial density across the different masses. We take the simulation with a mass of $10,000\,M_{\odot}$ a radius of $10\,\mathrm{pc}$ and $100,000$ SPH particles as fiducial with the others adjusted accordingly. We control the mass resolution rather than the number of SPH particles so that we can directly compare without considering the affect that mass resolution has on the sink particle masses. See full details in table \ref{tab:averagetable}. 

\begin{table*}
\centering
\begin{tabular}{rrrrrrrrrr}
\hline
Sim Mass $(M_{\odot})$ &  Sim Res &  $\alpha_{<500}$ &  $\alpha_{>500}$ &  Avg Time (Myr) &  Avg M\textsubscript{max} $(M_{\odot})$ &  M\textsubscript{max} $(M_{\odot})$ &  Sigma $(M_{\odot})$ &  Avg M\textsubscript{ecl} $(M_{\odot})$ & Required Sims \\
\hline
1000 &             1.0 &   0.651 &    0.173 &          5.00 &                  27.600 &          69.1 &  24.481 &                  37.160 &     36107909 \\
     1500 &             1.0 &   0.651 &    0.173 &          5.00 &                  55.080 &         114.6 &  32.290 &                  98.780 &         1287 \\
     2500 &             1.0 &   0.651 &    0.173 &          5.00 &                  94.140 &         168.2 &  34.915 &                 249.130 &           22 \\
     5000 &             1.0 &   0.651 &    0.173 &          5.00 &                 143.000 &         232.9 &  42.604 &                 694.650 &            2 \\
    10000 &             1.0 &   0.651 &    0.173 &          5.00 &                 163.600 &         242.2 &  40.025 &                1753.990 &            1 \\
    20000 &             1.0 &   0.651 &    0.173 &          5.00 &                 203.889 &         253.0 &  26.702 &                4129.922 &            1 \\
    40000 &             1.0 &   0.651 &    0.173 &          5.00 &                 223.300 &         346.4 &  58.606 &                9843.800 &            1 \\
     1000 &             2.0 &   0.431 &    0.106 &          5.00 &                  31.510 &          70.2 &  23.548 &                  47.650 &       384116 \\
     1500 &             2.0 &   0.431 &    0.106 &          5.00 &                  55.000 &          98.8 &  20.735 &                 110.100 &        38214 \\
     2500 &             2.0 &   0.431 &    0.106 &          5.00 &                  91.540 &         148.0 &  29.371 &                 260.130 &           12 \\
     5000 &             2.0 &   0.431 &    0.106 &          5.00 &                 114.520 &         168.2 &  32.906 &                 704.370 &            3 \\
    10000 &             2.0 &   0.431 &    0.106 &          5.00 &                 142.180 &         195.8 &  31.139 &                1758.270 &            1 \\
    20000 &             2.0 &   0.431 &    0.106 &          5.00 &                 146.240 &         186.3 &  26.914 &                4197.380 &            1 \\
    40000 &             2.0 &   0.431 &    0.106 &          4.93 &                 164.990 &         289.4 &  47.926 &                9041.080 &            1 \\
     1000 &             4.0 &   0.377 &    0.064 &          5.00 &                  29.730 &          50.7 &  14.853 &                  53.310 &       262396 \\
     1500 &             4.0 &   0.377 &    0.064 &          5.00 &                  48.790 &          89.2 &  19.651 &                 112.310 &           86 \\
     2500 &             4.0 &   0.377 &    0.064 &          5.00 &                  62.680 &         112.3 &  27.166 &                 266.700 &            6 \\
     5000 &             4.0 &   0.377 &    0.064 &          5.00 &                  89.310 &         138.4 &  25.435 &                 701.880 &            2 \\
    10000 &             4.0 &   0.377 &    0.064 &          5.00 &                  98.370 &         127.4 &  15.182 &                1750.450 &            1 \\
    20000 &             4.0 &   0.377 &    0.064 &          5.00 &                 102.400 &         130.5 &  14.611 &                4127.160 &            1 \\
    40000 &             4.0 &   0.377 &    0.064 &          4.90 &                 117.530 &         139.6 &  16.606 &                8477.050 &            1 \\
     1000 &             8.0 &   0.367 &    0.133 &          5.00 &                  24.930 &          49.9 &  11.731 &                  56.150 &        16002 \\
     1500 &             8.0 &   0.367 &    0.133 &          5.00 &                  35.390 &          59.9 &  14.758 &                 119.330 &           75 \\
     2500 &             8.0 &   0.367 &    0.133 &          5.00 &                  46.510 &          92.0 &  19.445 &                 270.970 &            6 \\
     5000 &             8.0 &   0.367 &    0.133 &          5.00 &                  61.820 &         108.2 &  20.857 &                 698.420 &            2 \\
    10000 &             8.0 &   0.367 &    0.133 &          4.98 &                  71.890 &         111.7 &  19.085 &                1659.890 &            1 \\
    20000 &             8.0 &   0.367 &    0.133 &          4.82 &                  72.540 &          92.7 &  10.708 &                3401.650 &            1 \\
    40000 &             8.0 &   0.367 &    0.133 &          4.58 &                  79.510 &          99.4 &  13.083 &                6520.150 &            1 \\
     1000 &            16.0 &   0.437 &    0.209 &          5.00 &                  18.640 &          36.1 &   9.401 &                  57.820 &         1990 \\
     1500 &            16.0 &   0.437 &    0.209 &          5.00 &                  23.830 &          48.6 &  11.145 &                 120.450 &           74 \\
     2500 &            16.0 &   0.437 &    0.209 &          5.00 &                  36.900 &          61.8 &  13.904 &                 263.140 &            4 \\
     5000 &            16.0 &   0.437 &    0.209 &          5.00 &                  43.270 &          51.4 &   7.574 &                 683.910 &            4 \\
    10000 &            16.0 &   0.437 &    0.209 &          4.91 &                  51.350 &          70.1 &  12.500 &                1585.600 &            1 \\
    20000 &            16.0 &   0.437 &    0.209 &          4.30 &                  47.920 &          60.9 &  11.209 &                2299.410 &            1 \\
    40000 &            16.0 &   0.437 &    0.209 &          3.92 &                  48.970 &          85.9 &  13.800 &                3665.310 &            1 \\
     1000 &            32.0 &   0.313 &    0.113 &          5.00 &                  12.840 &          24.6 &   6.926 &                  57.650 &           73 \\
     1500 &            32.0 &   0.313 &    0.113 &          5.00 &                  18.460 &          39.5 &   7.881 &                 120.930 &            7 \\
     2500 &            32.0 &   0.313 &    0.113 &          4.90 &                  20.470 &          27.9 &   4.468 &                 229.040 &           29 \\
     5000 &            32.0 &   0.313 &    0.113 &          4.81 &                  25.170 &          31.2 &   4.508 &                 514.930 &            4 \\
    10000 &            32.0 &   0.313 &    0.113 &          4.66 &                  29.880 &          41.5 &   7.448 &                1152.830 &            1 \\
    20000 &            32.0 &   0.313 &    0.113 &          4.26 &                  30.800 &          41.1 &   5.641 &                2156.540 &            1 \\
    40000 &            32.0 &   0.313 &    0.113 &          3.48 &                  28.550 &          37.3 &   6.535 &                2274.490 &            1 \\
\hline
\end{tabular}
\caption{Table contains simulation specifications and results averaged according to simulation mass and resolution. $\alpha_{<500}$ and $\alpha_{>500}$ are power law slopes corresponding to fits to most-massive star masses over clusters masses below and above $500\,M_{\odot}$ respectively. 'Required Sims' is the number of simulations we would need to run before we would expect to find a star with the average maximum mass formed in the $10,000\,M_{\odot}$ simulations at the same resolution.}
\addtocounter{table}{0}
\label{tab:averagetable}
\end{table*}

For the resolution variation we repeatedly double the number of SPH particles until the simulations become prohibitively computationally expensive, a resolution indicator of 1 means that there are 10 particles per solar mass. The sink particle formation criteria are also adjusted adhering to SPH resolution, where we adjust the sink particle critical density such that a sink particle will form from at least 100 SPH particles \citep{Bate1997sphresolution}. Therefore at 1 resolution the minimum mass of a sink particle is $10\,M_{\odot}$. This set of simulations also serves to bring clarity to the massive star vs. small association ambiguity found in our more massive simulations (e.g. the $40,000\,M_{\odot}$ res = 1 simulation) described above by allowing a small association to be resolved into individual sink particles. 

For all of the simulation specifications above we perform $10$ simulations varying the random seed used to create the turbulent field. We do this to obtain a statistically significant result by reducing the errors inherent in a simulation of a chaotic system \citep{Jaffa2022}.

We investigate the IMF of the sink particles in order to analyse key properties of the sink population. We look at the maximum mass of a sink particle achieved in order to study how the total cloud mass affects the maximum sink particle mass. The minimum mass found in the sink population will be limited by the mass resolution \citep{Bate1997sphresolution}. Therefore the minimum mass possible will scale inversely with the number of SPH particles. The presence of a power law slope in the high mass region of the IMF serves as a sanity check that we have a realistic distribution of stellar masses.

We investigate the masses of the most massive sink particles for each simulation as well as the mean and standard deviation. This shows us the maximum mass for each simulation, the spread for a given mass, and whether the maximum masses had converged. We use the standard deviations to estimate how many simulations at a given mass we would need to perform to be likely to form a star as massive as we find in our $10,000\,M_{\odot}$ at the same respective resolution. We choose $10,000\,M_{\odot}$ for the comparison as the higher masses often don't run for the full duration due to numerical problems and therefore their stellar and cluster masses are understated.   
We also look at how the maximum mass changes with mass resolution by plotting the relation between the average cluster mass and average maximum stellar mass at multiple resolutions. This allows us to see the effect that mass resolution has on the star formation and demonstrates the importance of comparing results at the same mass resolution.

To examine the possibility that our clusters could be considered to be multiple smaller clusters we use a friend-finding algorithm \citep{davis1985Groups}. This algorithm separates our clusters into groups according to a "linking length", a group is then made of any sink particles that can be joined by no more than this length.

\section{Results and Analysis}
\label{sec:results}
\begin{figure}
	
    \includegraphics[width=0.5\textwidth]{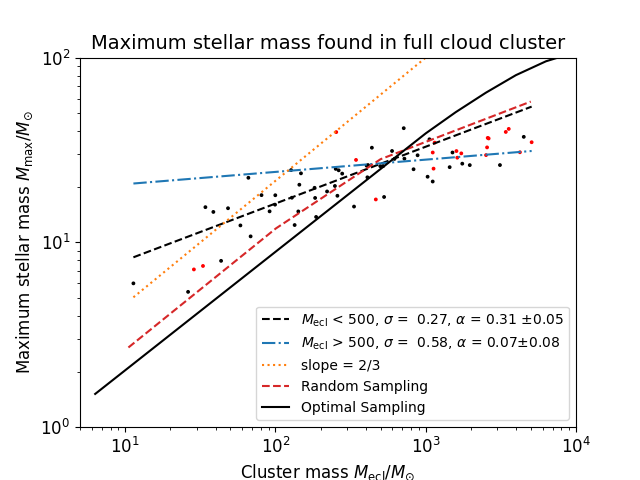}
    \includegraphics[width=0.5\textwidth]{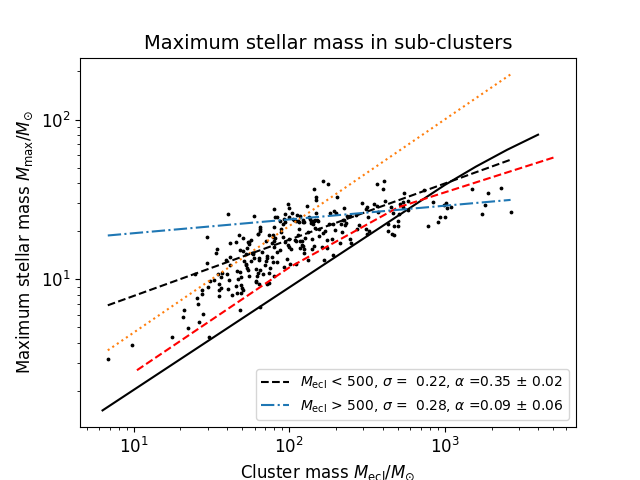}
    \caption{\textbf{Top}: Maximum star mass against total stellar mass in all of the highest resolution (32 resolution) simulations. Random sampling plotted as red dashed line, created using mass constrained random sampling from a Kroupa IMF up to $150\,M_{\odot}$. Optimal sampling plotted as a solid black line from \citet{Weidner2010Mmass_Mecl}. Dotted line shows an exponent of $2/3$ \citep{bonnell2004}. Simulations that did not reach the full runtime are plotted as red dots (at least $90\%$ of the nominal runtime), black dots represent completed simulations. Black dashed and blue dot-dashed lines show power-law fit lines to our data below and above $M_{\mathrm{ecl}} = 500\,M_{\odot}$, respectively. In the legend, we give the respective power-law index ($\alpha$) with uncertainties (one standard deviation) from the fit as well as a measure of the scatter of the data points around the fit line ($\sigma$, also one standard deviation). \textbf{Bottom}: The same plot as the top panel but each cluster is split into groups using a friend-finding algorithm. Each group is then treated as its own cluster.} 

    \label{fig:allmax}
\end{figure}

At our highest resolution, the mass of the most massive stars in our simulations correlates with the mass of the cluster formed (Fig \ref{fig:allmax}). This behaviour is expected from both random sampling and optimal sampling, as can be seen in Fig~\ref{fig:allmax}: The red dashed line shows the mean maximum star mass expected for random sampling, and the solid black line shows the exact value for the most massive star for a given cluster mass for optimal sampling. While the two lines are similar, we expect the most massive star masses to be scattered around the lines for random sampling, but exactly on the line for optimal sampling if the respective sampling method was a faithful description of our simulations. Our simulations clearly show a scatter of most massive star masses for any given cluster mass, which is in general agreement with the random sampling concept. It is possible that our clusters could be considered to be comprised of multiple smaller clusters. To address this we use a friend-finding algorithm to group the sink particles according to a 'linking length'. We find that the clusters have no preffered scale with the grouping changing smoothly with the linking length. This makes sense for a cluster formed from a cloud with decaying turbulence.

\begin{figure}
	
    \includegraphics[width=0.5\textwidth]{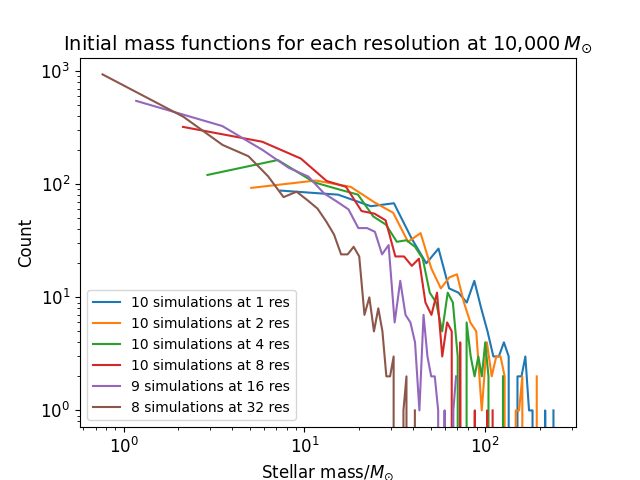}
    \caption{The initial mass function of the $10,000\,M_{\odot}$ simulations for the range of resolutions as labelled.}
    \label{fig:IMFstack}
\end{figure}

\begin{figure}
	\includegraphics[width=0.5\textwidth]{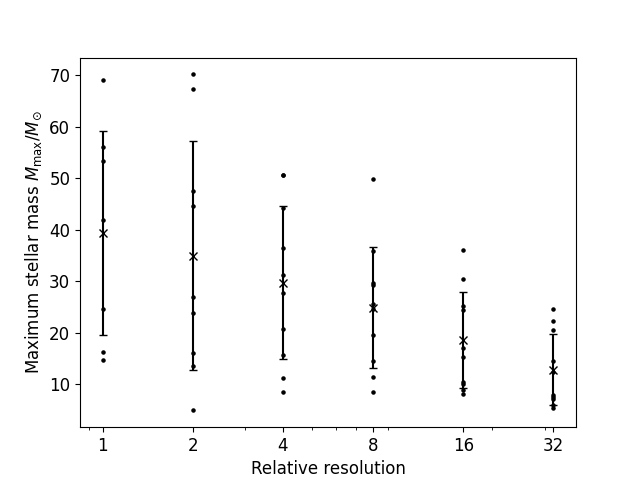}
    \includegraphics[width=0.5\textwidth]{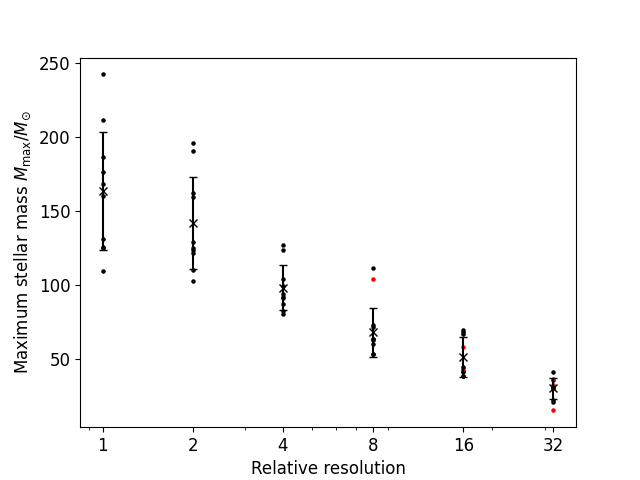}
    
    \includegraphics[width=0.5\textwidth]{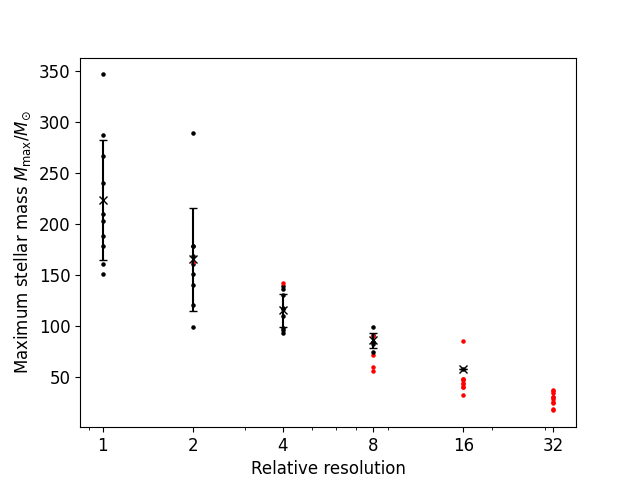}
    
    \caption{Distribution of maximum sink particle mass as we increase the mass resolution for the \textbf{Top}: $1,000\,M_{\odot}$, \textbf{Middle}: $10,000\,M_{\odot}$, and \textbf{Bottom}: $40,000\,M_{\odot}$ simulations. The maximum sink mass decreases with resolution as expected, it increases with increased cloud mass between the $1,000\,M_{\odot}$ and the $10,000\,M_{\odot}$ plots. There is less difference between the $10,000\,M_{\odot}$ and $40,000\,M_{\odot}$. The red dots indicate simulations that did not run to completion, these are removed from calculations of the error bars.}
    \label{fig:maxressink}
\end{figure}

\begin{figure}
	
    \includegraphics[width=0.5\textwidth]{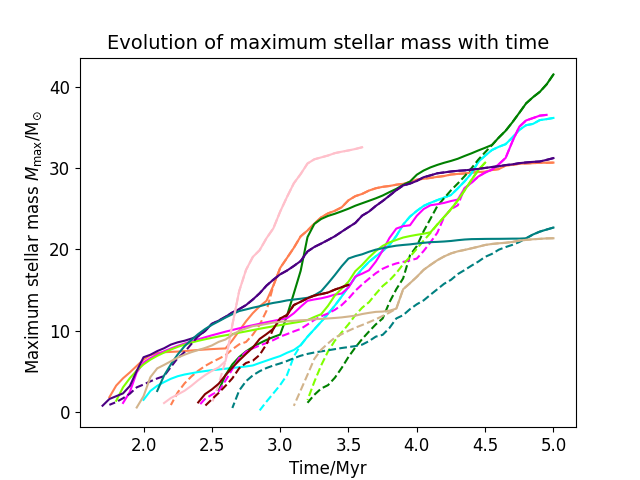}
    \caption{Mass evolution of the final most massive star (dashed line) as well as the highest mass at each time (solid line) for each random seed with 32 resolution and initial cloud mass $10,000\,M_{\odot}$. }
    \label{fig:massevo}
\end{figure}

\begin{figure*}
	
    \includegraphics[width=0.4\textwidth]{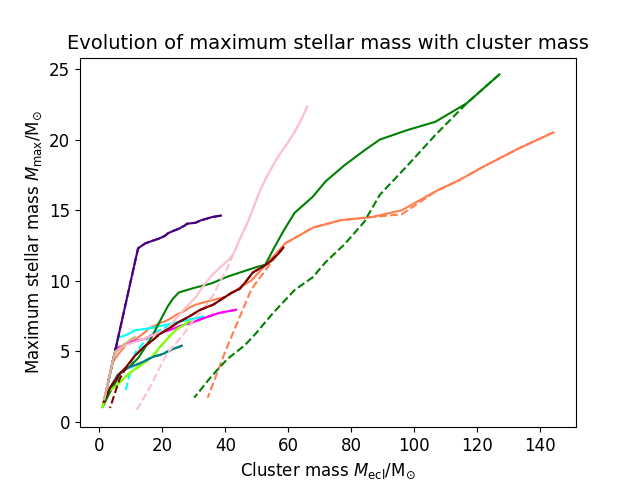}
    \includegraphics[width=0.4\textwidth]{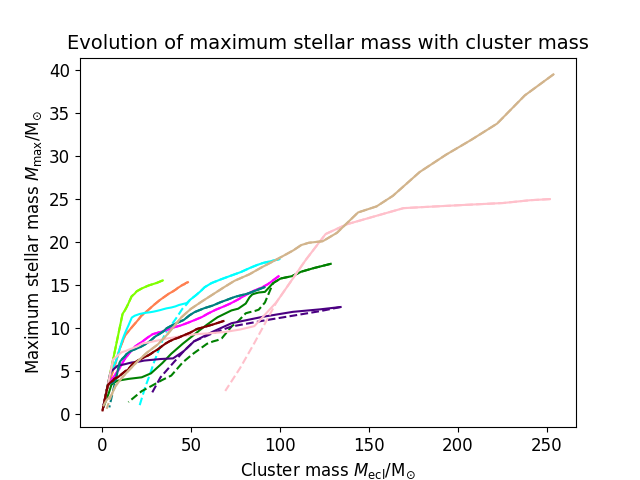}
    \includegraphics[width=0.4\textwidth]{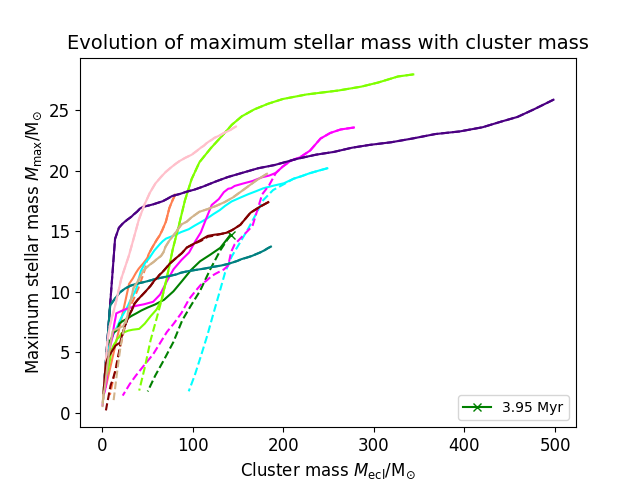}
    \includegraphics[width=0.4\textwidth]{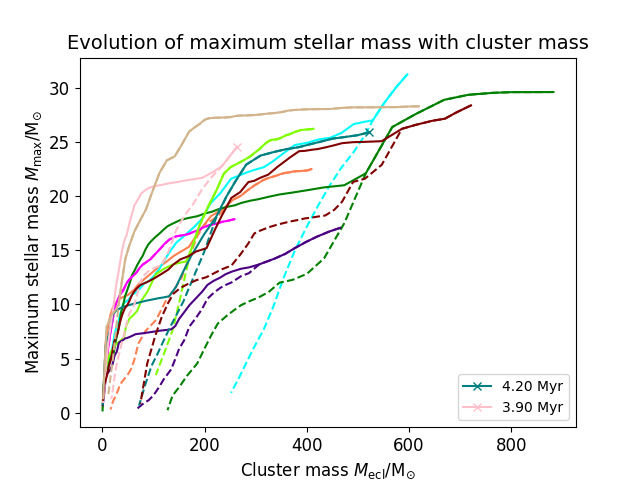}
    \includegraphics[width=0.4\textwidth]{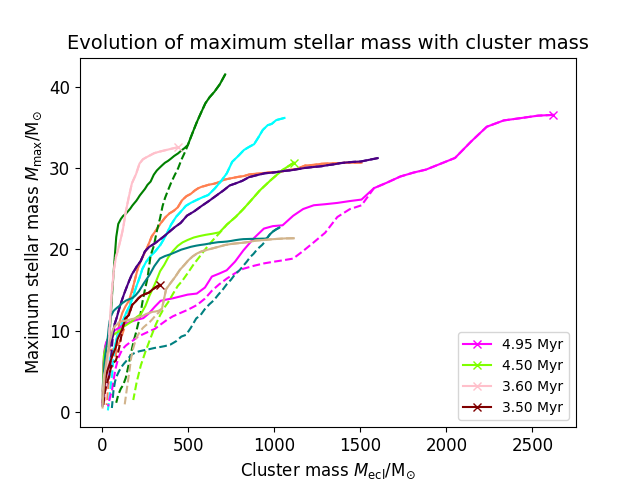}
    \includegraphics[width=0.4\textwidth]{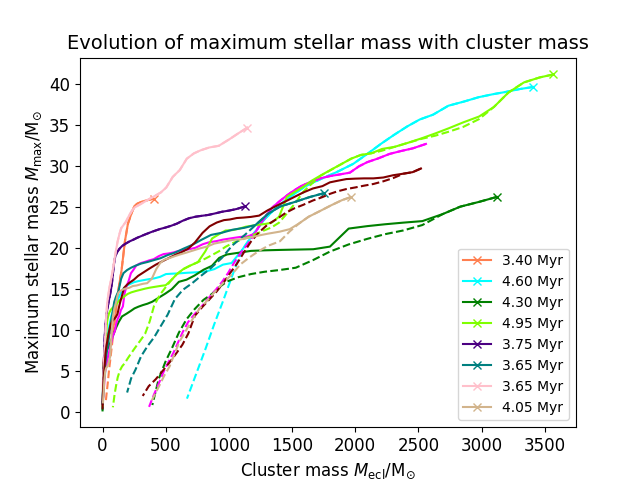}
    \includegraphics[width=0.4\textwidth]{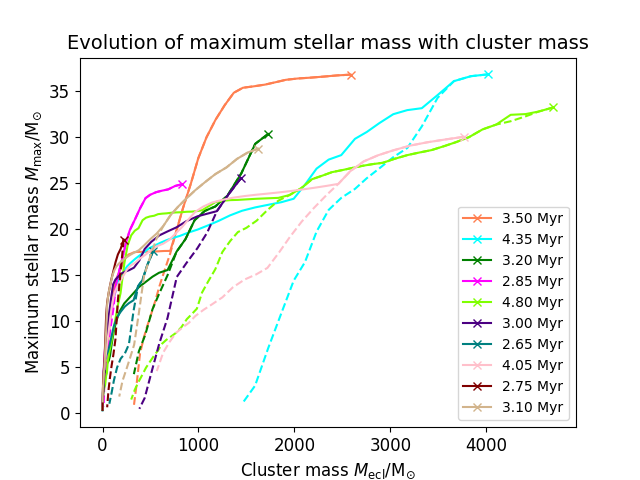}

    \caption{The evolution of the mass of the most massive star against the cluster mass. Solid line represents the most massive star at that time while the dashed line tracks the star that will end up as the most massive. Resolution and simulation mass labelled. A '$\times$' indicates that that individual simulation did not reach 5\,Myr, the end times are labelled in these cases.}
    \label{fig:massclevo}
  \end{figure*}
\begin{figure}
	
    \includegraphics[width=0.5\textwidth]{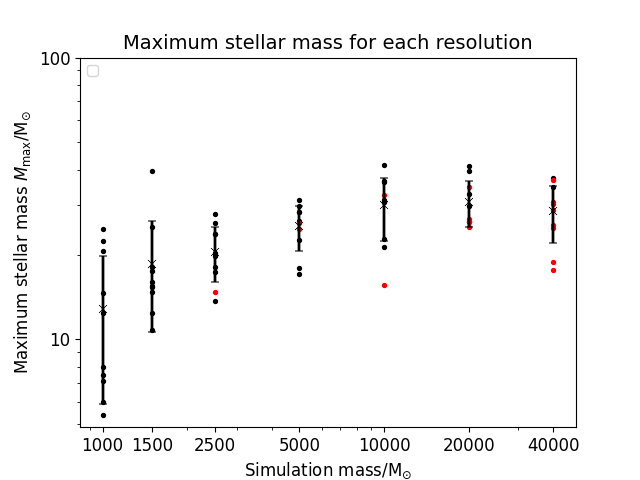}
    \caption{Distribution of maximum star masses found in each of the simulations we examine at their maximum mass resolution. Red dots indicate a simulation that didn't run for at least $4.5\,Myr$}
    \label{fig:maxsink}
\end{figure}

Fig \ref{fig:allmax} (bottom) shows the results for an example linking length of 1 pc changes are subtle. Overall, the result is very similar to taking all stars in a given simulation as a cluster (Fig \ref{fig:allmax}, top). For full-simulation clusters and below a cluster mass $M_\mathrm{ecl}$ of 500 $M_\odot$, we find for the mass of the most massive star, $M_\mathrm{max}\propto M_\mathrm{ecl}^\alpha$ with $\alpha=0.31\pm0.05$. For  $M_\mathrm{ecl}>500M\odot$ $\alpha$ flattens to $0.07\pm0.08$. The values for $\alpha$ are consistent within uncertainties for the clusters defined via the group-finding algorithm. The scatter around these lines is similar below 500 $M_\odot$. Above this limit, the scatter is somewhat reduced for the clusters defined by the group-finding algorythm.

Improving the mass resolution of the simulations alters the IMF (Fig \ref{fig:IMFstack}) in a number of ways. First we can see that the low mass end shifts to lower masses with improved resolution as expected, we also see a shift to lower masses in the rest of the IMF following the low mass end, finally we see that the high mass end peaks at lower masses as we increase the resolution.

These results might suggest that there is a critical cloud mass required to form massive stars. In our simulations this critical mass is resolution dependent. In our low resolution simulations we require a cloud mass of $5,000\,M_{\odot}$ to form a sink particle above $100\,M_{\odot}$. At our highest resolution we find a critical cloud mass of $10,000\,M_{\odot}$ required to produce a sink mass of $40\,M_{\odot}$ (Fig. \ref{fig:maxsink}). In Table \ref{tab:averagetable} we show the level of outlier required for each simulation setup to produce the average maximum mass found in the $10,000\,M_{\odot}$ simulation of the same resolution. We see that, at the highest resolution, for a $1,000\,M_{\odot}$ simulation a $2.46\sigma$ outlier is required to form the average mass found in our $10,000\,M_{\odot}$ simulations. This corresponds to needing $72$ simulations of $1,000\,M_{\odot}$ to form the $10,000\,M_{\odot}$ average. Since the mass ratio between the two setups is only a factor of $10$, this means that stars towards the upper mass end of the $10,000\,M_{\odot}$ cloud form much less frequently in our $1,000\,M_{\odot}$ clouds than what would be expected from random sampling from the IMF from our $10,000$ solar mass clouds.

This decrease in the maximum mass is better seen by looking at the distribution of maximum masses directly (Fig. \ref{fig:maxressink}). This clearly shows the masses decreasing with each increase in resolution, this is expected as the better resolution allows us to resolve the larger dense regions into multiple sink particles rather than fewer large sink particles.

The higher mass ($40,000\,M_{\odot}$) resolution study shows the same trends (Fig. \ref{fig:maxressink}). This also allows us to see that the very high mass (>$200\,M_{\odot}$) sink particles seen in Fig~\ref{fig:maxressink} are most likely representing an unresolved group. We see that the maximum stellar masses drop as we increase resolution, as the potential for sink particles to represent many individual stars is reduced, so is the variation on the sink mass reduced to the variation of the mass of an individual star. Resolution is further examined in the Appendices.

From the mass evolution of the most massive star over time (Fig \ref{fig:massevo}) we see that the final most massive star has been the most massive throughout the simulation time in approximately half of the simulations. In the majority of cases the star appears to still be growing, however, this growth is significantly reduced in the lower mass simulations except for a few cases. In the cases where we see a late forming star become the most massive, it is often the case that the previous most massive is only growing slowly due to it having exhausted its local reservoir. The resolution dependence of the maximum stellar mass is not sufficiently explained by lower formation mass (due to aforementioned resolution criteria), instead it is due to complex structure in the gas leading to erratic bursts of higher accretion. 

Fig. \ref{fig:massclevo} shows the mass evolution of the most massive star with the cluster mass. In the lower mass simulations we see that the most massive stars are found in the most massive clusters. This is less the case in the higher mass simulations where the majority of the clusters reach >$500\,M_{\odot}$. The dominant impression of leveling off in these plots means that while the most massive star quickly exhausts its immediate environment, other regions of the cluster keep growing steadily at late times. There is an interesting case at a cloud mass of $1500\,M_{\odot}$, where we find one simulation where the most massive star keeps growing strongly and in proportion with the rest of the cluster, until it reaches $40\,M_{\odot}$ at a cluster mass of only $250\,M_{\odot}$. This illustrates that occasionally, very high-mass stars can form in relatively low-mass clusters in our simulations.

The behaviour seen with mass evolution against cluster mass (Fig \ref{fig:massclevo}) is split at M\textsubscript{ecl} \textasciitilde 500\,$M_{\odot}$. Below this we see a consistent positive correlation between cluster mass and star mass, above this the maximum star mass depends more on the cloud mass, this becomes more prominent if we factor in the shorter average runtime of the higher mass simulations. While resolution does affect the sink masses significantly, due to the minimum possible mass and ability to resolve large sinks into multiple smaller sinks, this affect is fairly consistent across the different masses (Fig. \ref{fig:maxressink} and Appendix \ref{appendix:a} Fig. \ref{fig:totalstack}).
\section{Discussion}

\label{sec:discussion}
The suite of simulations presented here consistently displays certain trends found when varying both the cloud mass and the mass resolution. We see that an increase in the cloud mass invariably leads to higher mass sink particles being formed, this trend is subtle at higher cloud masses but is very obvious up to the intermediate cloud masses ($\sim2,000\,M_{\odot}$). While the more extreme of these massive sink particles can be explained as unresolved groups (as indicated by their absence at higher resolution) this trend is still apparent when comparing the highest mass resolution simulations (Fig. \ref{fig:maxressink}). However, very rarely, we also find very high-mass stars in comparatively low-mass clusters (see $1500\,M_{\odot}$ in Fig \ref{fig:massclevo}).

This would suggest that the very high mass stars found in apparent isolation \citep[e.g.][]{bestenlehner2011VLTFLAMES,Bressert2012VLT-FLAMES,Oskinova2013Isolation} either didn't form in isolation or they formed under extreme circumstances not included in our simulations.

We see in Fig. \ref{fig:allmax} that the maximum star mass increases with cluster mass a trend that appears, in principle, consistent with the findings presented by \cite{Weidner2013MmaxMecl2} (see their Fig. 1). We also see significant scatter in the maximum mass for a given cloud mass, this spread decreases with increased cloud mass. This suggests that the spread \cite{Weidner2013MmaxMecl2} find may be less due to observational uncertainty and more due to actual variation than they suggest. The findings shown in our Fig. \ref{fig:maxressink} at first appear inconsistent with the idea of purely stochastic star formation. At all resolutions we see higher mass stars in higher mass clusters. There is, however, a scatter involved, and had we carried out even more simulations, we might have found even higher most-massive-star values. To quantify this we compared each simulation (cloud) mass to the $10,000\,M_{\odot}$ simulations at the same respective resolution. We calculate how many low mass clusters on average are required to form a star of the same mass. We see that for low mass clusters to form a star with the average maximum mass of a high mass cluster (e.g. $10,000\,M_{\odot}$) would also form more cluster mass than is required on average for the $10,000\,M_{\odot}$ simulation, for example $72$ simulations at $1,000\,M_{\odot}$ would form over $4,000\,M_{\odot}$ of cluster mass (almost 4 times more than the single $10,000\,M_{\odot}$ simulation). This is in clear disagreement with the expectations of purely random sampling where the same total cluster mass should produce the same massive stars on average. We see the largest change in maximum mass seen between cloud masses $1,000\,M_{\odot}$ and $5,000\,M_{\odot}$ after which the change is slighter as the mass increases. The maximum mass averaged over each repeat simulation also increases up to $10,000\,M_{\odot}$. However it remains fairly consistent after that point.
From our look at potential sub-clusters we find that in both cases the most-massive-star mass against cluster mass becomes shallower at $500\,M_{\odot}$ but the power-law slopes are consistent within the uncertainties. The scatter decreases in both the high-mass and low-mass groups, though the change is greater at the high-mass end. This makes sense as the maximum mass will remain the same while the sub-cluster mass will be lower than the combined cluster.

We observe that our most-massive-star mass with cluster-mass differs in detail to both the expectations of random sampling and optimal sampling (compare Fig. \ref{fig:allmax}). At high cluster masses, our most-massive-star masses are below the expectaions, whereas at low cluster masses we find higher star masses than expected. Yet, our quantitative analysis in Table \ref{tab:averagetable} shows that in order for all our clusters to be randomly sampled from the same mass function we would need to see even higher mass stars in our low mass clusters. This apparent contradiction demonstrates that our clusters are not randomly sampled from the Kroupa IMF. Looking at Fig. \ref{fig:IMFstack} we see that as we increase resolution our mass functions get steeper, approaching the Kroupa IMF. At the same time, at higher resolution we need fewer low mass clusters to get a most-massive-star as massive in a high cluster-mass simulation, in better agreement with random sampling of a uniform IMF. This gives reason for hope that in the future even higher resolution simulations will agree with observations.

Fig. \ref{fig:massevo} shows the mass evolution against time for the $10,000\,M_{\odot}$ mass clouds at our highest resolution. We see that in all of the simulations the most massive star continues to grow, though the rate of growth varies with random seed. This is likely due to variation in the rate of accretion onto the system from the surrounding cloud. After their initial burst of growth the star growth slows, either almost flattening out or becoming 'clumpy'. The stars that flatten out are often overtaken by late forming stars with more rapid growth. This behaviour is possibly due to gas supply with growth slowing when the stars birth clump is depleted and further growth relying on gas being funnelled into the system. This could either be slow and steady or clumpy resulting in the erratic rate of growth we see in some of the stars.

To look at the effects that the physics missing from our simulations may have had, we compare our results to those from \cite{Guszejnov2022}. They find that adding feedback to their simulations decreases the maximum mass found. Their resulting most massive stars in their "M2e4\_C\_M\_J" simulations are of similar mass to our higher mass simulations. \citet{Grudic2023} perform a similar study to ours simulating many lower mass clusters with feedback. They too find that a single larger cluster will produce more massive stars than many lower mass clouds.

\section{Summary and Conclusions}
In this paper we analysed the statistical properties of the massive-star populations that form for molecular clouds with a range of different masses. The goal was to see if there was a significant difference for star formation in low mass star clusters and high mass star clusters. We consider two extremes: first that star formation is purely stochastic and a given combined stellar mass will be comprised of stars of statistically the same masses independently of whether they were in a single very massive cluster or many low mass clusters (random sampling). Secondly that star formation is completely deterministic and that for a given cluster mass there is a set maximum stellar mass \citep[\textit{Optimal Sampling, }][]{kroupa2013}.

Our simulations do not entirely agree with either of the above options. We see significant scatter in the maximum star mass produced from the same initial conditions. This rules out deterministic star formation, though we note that various forms of feedback missing from our simulations could potentially inhibit accretion once a star reaches a certain mass and thus reduce the scatter. We also find a significant trend, most noticeable at lower masses, between cluster mass and maximum star mass. We also find a critical mass requirement to form stars above a certain mass ($40\,M_{\odot}$ stars are not found below a cluster mass of $500\,M_{\odot}$ in our highest resolution simulations). These combined show that in our simulations star formation cannot be purely stochastic. 

From the calculated standard deviations of the stellar masses for each cloud mass and simulation resolution we also see that the probability of a low mass cloud forming a star as massive as are often formed from high mass clouds is sufficiently low so that we would need to form much more cluster mass before we would expect to see a star of similar mass to the higher mass simulation (4 times the cluster mass from the simulations with cloud masses of $1,000\,M_{\odot}$ vs. $10,000\,M_{\odot}$, respectively, compare above). Therefore, our low mass clusters are not randomly sampled from our high mass clusters' distributions. This further disagrees with purely stochastic star formation which predicts the massive star population to be consistent with total cluster mass. Compared to both random and optimal sampling based on observed initial mass functions, our low mass clusters still form too many massive stars. On the other hand our high mass clusters do not reach observed massive star masses.
We note that the required number of low-mass-cluster simulations to yield a most-massive star as massive as in a higher-mass simulation decreases with increased resolution. It is thus possible that at very high resolution we may see agreement with random sampling.

We see from the evolution for maximum stellar mass with cluster mass (Fig \ref{fig:massclevo}) that the paths the most massive stars take varies significantly. While sometimes an early forming star will steadily accrete and end up as the most massive star in the cluster, it is also common for the early forming star's growth to slow significantly and for a late forming star to overtake with more rapid accretion. This demonstrates that the star that eventually becomes the most massive star is not "linearly" predetermined by the initial conditions, but emerges dynamically by the non-linear behaviour of the system. 

\label{sec:conclusions}

\section*{Acknowledgements}
SJ acknowledges support from the STFC grant ST/R00905/1. JDS
acknowledges a studentship from the Science and Technology Facil-
ities Council (STFC) (ST/T506126/1).
We thank the reviewer for their helpful suggestions and constructive criticism.

%%%%%%%%%%%%%%%%%%%%%%%%%%%%%%%%%%%%%%%%%%%%%%%%%%
\section*{Data Availability}
Data and full running instructions available on request: j.smith49@herts.ac.uk
 
%The inclusion of a Data Availability Statement is a requirement for articles published in MNRAS. Data Availability Statements provide a standardised format for readers to understand the availability of data underlying the research results described in the article. The statement may refer to original data generated in the course of the study or to third-party data analysed in the article. The statement should describe and provide means of access, where possible, by linking to the data or providing the required accession numbers for the relevant databases or DOIs.



\bibliographystyle{mnras}
\bibliography{References} % if your bibtex file is called example.bib

%\begin{thebibliography}{99}
%\end{thebibliography}

%%%%%%%%%%%%%%%%%%%%%%%%%%%%%%%%%%%%%%%%%%%%%%%%%%

%%%%%%%%%%%%%%%%% APPENDICES %%%%%%%%%%%%%%%%%%%%%

\appendix

\section{Resolution study}
\label{appendix:a}
\begin{figure}
\includegraphics[width=0.5\textwidth]{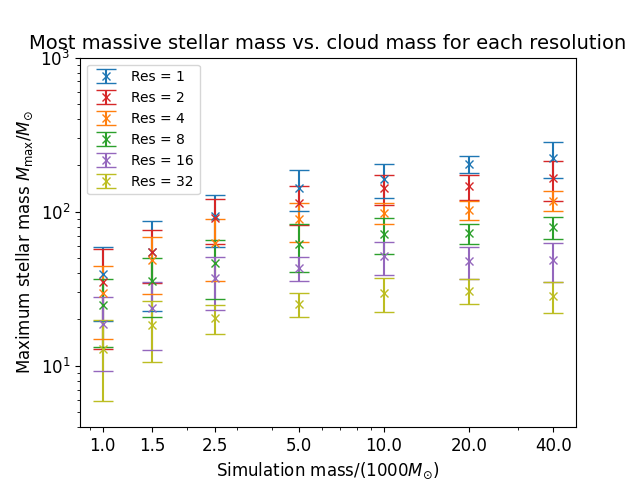}
\caption{Average maximum mass for each simulation cloud mass at each resolution. One standard deviation plotted as error bars, See legend for correspondence between colours and resolution. The maximum star mass decreases monotonically with increasing resolution for all simulated clouds.
For all resolutions, the maximum star mass increases up to 5,000 Msun, and then stays roughly constant with simulation mass.}
\label{fig:totalstack}
\end{figure}

We show mean most massive star masses with one-sigma variation between different realisations of the same simulation mass over the total mass in the simulations in Fig. \ref{fig:totalstack}. We do this to easily display the differences both mass and resolution have on the maximum mass. We see that maximum mass is well correlated with simulation mass and anti-correlated with simulation resolution.
%%%%%%%%%%%%%%%%%%%%%%%%%%%%%%%%%%%%%%%%%%%%%%%%%%

\begin{figure}
    \includegraphics[width=0.5\textwidth]{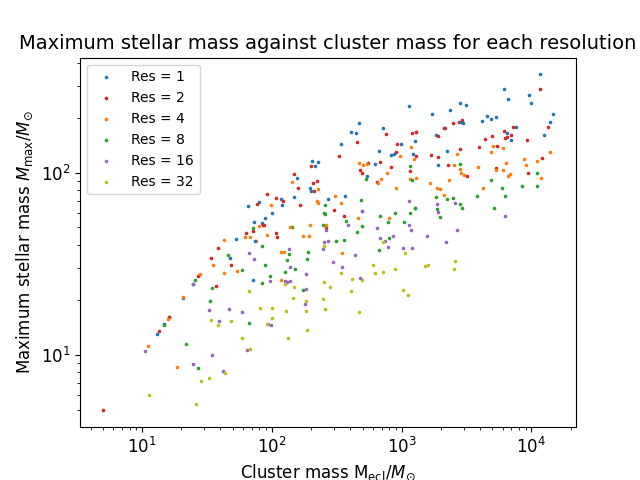}
    \caption{Maximum stellar mass against cluster mass for all simulations, with the resolutions indicated by colour as labelled. A stellar mass limit is reached for a given cluster mass which decreases with increasing resolution.}
    \label{fig:allmaxes}
\end{figure}

We plot the maximum sink particle mass over cluster mass for every simulation (all that formed sink particles) and these are shown in Fig. \ref{fig:allmaxes}. We see again that higher resolution consistently reduces maximum stellar mass and that higher simulation cloud mass results in higher maximum stellar mass, the cloud mass dependence is stronger at lower resolutions but is still present at our highest resolution with maximum masses increasing more steeply until about $600\,M_{\odot}$ where we see the trend flatten.

% Don't change these lines
\bsp	% typesetting comment
\label{lastpage}
\end{document}